# Comparing the binding interactions in the receptor binding domains of SARS-CoV-2 and SARS-CoV

Muhamed Amin[†,‡] *, Mariam K. Sorour[§], Amal Kasry[§]

[†] Department of Sciences, University College Groningen, University of Groningen, Hoendiepskade 23/24, 9718 BG Groningen, The Netherlands
[‡] Center for Free-Electron Laser Science, Deutsches Elektronen-Synchrotron DESY, Notkestrasse 85, 22607 Hamburg, Germany
[§] Nanotechnology Research Centre (NTRC), the British University in Egypt (BUE), El Sherouk City, Suez Desert Road, Cairo 1183.

m.a.a.amin@rug.nl

COVID-19, since emerged in Wuhan, China, has been a major concern due to its high infection rate, leaving more than one million infected people around the world. Huge number of studies tried to reveal the structure of the SARS-CoV-2 compared to the SARS-CoV-1, in order to suppress this high infection rate. Some of these studies showed that the mutations in the SARS-CoV-1 Spike protein might be responsible for its higher affinity to the ACE2 human cell receptor. In this work, we used molecular dynamics simulations and Monte Carlo sampling to compare the binding affinities of the spike proteins of SARS-CoV and SARS-CoV-2 to the ACE2. We found that the SARS-CoV-2 binds to ACE2 stronger than SARS-CoV by 7 kcal/mol, due to enhanced electrostatic interactions. The major contributions to the electrostatic binding energies are resulting from the salt-bridges formed between R426 and ACE2-E329 in case of SARS-CoV and K417 and ACE2-D30 for SARS-CoV2. In addition, there is no significant contribution from a single mutant to the binding energies. However, these mutations induce sophisticated structural changes that enhance the binding energies. Our results also indicate that the SARS-CoV-2 is unlikely a lab engineered virus.

**TOC GRAPHICS**

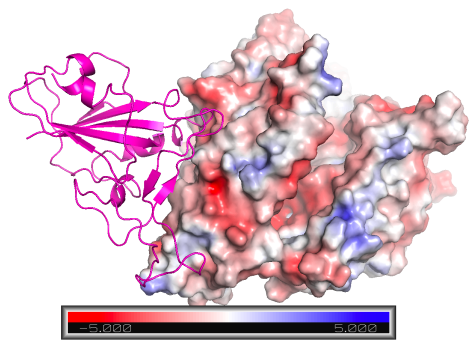

**KEYWORDS**
SARS-CoV, SARS-CoV-2, binding domains, Molecular Dynamics, Monte Carlo

A highly infectious strain of the SARS-CoV known as SARS-CoV-2 emerged in Wuhan China and was first reported on the 31st of December 2019, according to the World Health Organization (WHO)[1]. This strain has already resulted in more than 1 million infections and over 60,000 deaths all over the world. Previous studies conducted on the first SARS-CoV strain showed that the most likely mechanisms by which the virus interacted with a cell, were mediated by the receptor binding domain (RBD) on the S protein, which binds the human angiotensin-converting enzyme 2 (ACE2)[2].

Similarly, SARS-CoV-2 was proposed to garner entry to the cells through the RBD.[3,4,5] However, there are several mutations identified in the RBD, which include: $V404 \rightarrow K417$, $R426 \rightarrow N439$, $Y442 \rightarrow L455$, $L443 \rightarrow F456$, $L460 \rightarrow F473$, $L472 \rightarrow F486$, $L479 \rightarrow Q493$, $D480 \rightarrow S456$, $Y484 \rightarrow Q498$, $T487 \rightarrow N501$, according to the latest resolved structure using Cryo-electron microscopy.[5] Furthermore, another study using comparative genomics identified only 5 mutations and stated that RBD does not contain insertion or deletion.[6] The $V404 \rightarrow K417$ mutation balances the $R426 \rightarrow N439$ since they include the addition and removal of positively charged amino acids respectively. However, the $D480 \rightarrow S456$ mutation removes a negatively charged amino acid and is expected to make the electrostatic surface potential of SARS-CoV-2 more positive.

In this work, we study the binding energies between the ACE2 receptor and the Spike protein of SARS-CoV and SARS-CoV-2 viruses using Monte Carlo (MC) sampling and Molecular Dynamics (MD) simulations.

To study the interactions between the SARS-CoV and the ACE2, the crystal structure 2AJF,[1] which includes the receptor binding domain RBD is MD optimized using openMM software.[9-14]. Then, the optimized structure is inputted to MCCE (Multi Conformer Continuum Electrostatic)

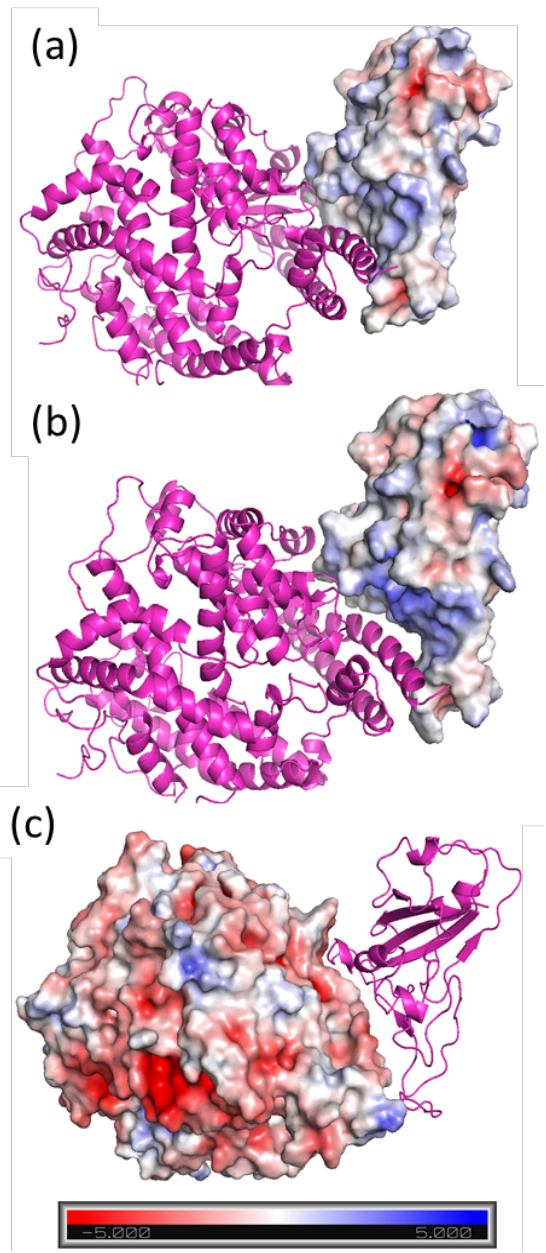

**Figure 1.** The electrostatic potential maps in kT/e of (a) the SARS-CoV, (b) SARS-CoV-2 and C) the ACE2.

software[16] to sample the protonation states of the amino acids using Monte Carlo. The electrostatic interactions between the different amino acids' conformers are calculated by solving Possion Boltzmann equation using

DELPHI.[17] The generated conformers' occupancies based on Boltzmann distributions are used to calculate the electrostatic and van der Waals interactions between the amino acids in the SARS-CoV and the ACE2.

However, to study the interactions between the SARS-CoV-2 and the ACE2, we constructed the following mutants (V404K, R426N, Y442L, L443F, F460Y, L472F, N479Q, D480S, Y484Q, T487N) based on the cryoEM structure PDB ID 6M17.[7] The mutations were performed by replacing the sidechains in the native structures with the proper sidechains of the mutants using MCCE. Several confirmations of the sidechains were created to void van der Waals clashes. The sidechain conformers with the highest Boltzmann occupancies were then MD optimized. The MCCE was then used to calculate the binding energies based on the optimized structures.

The electrostatic potential maps of SARS-CoV, SARS-CoV-2 and ACE2 were calculated using Adaptive Poisson-Boltzmann Solver (APBS)[16] (**Figure1.** a, b, and c). The ACE2 exhibit a negative electrostatic potential at the RBD (**Figure1.** c), while both SARS-CoV and SARS-CoV-2 show positive potentials. However, the potential observed for SARS-CoV-2 is more positive, which will result in a greater electrostatic attraction between ACE2 and SARS-CoV-2. This could be explained by the replacement of the negatively charged amino acid Asp by a neutral Ser (D480S).

The optimized structure of SARS-CoV shows salt-bridges formed between SARS-CoV-R426 and ACE2-E329, which dominates the electrostatic interactions with the ACE2. This salt-bridge is replaced by another salt-bridge formed between CoV-2-K404 and ACE2-D30 in SARS-CoV-2 (**Figure 2**). However, the CoV2-K404/ACE2-D30 salt-bridge has more favorable electrostatic. attractions by 1.4 Kcal/mol (**Table 1**). The total electrostatic interactions between the SARS-CoV-2 and the ACE2 are stronger by 3 Kcal/mol than SARS-CoV (**Table S1**). The contribution from the mutants themselves to the electrostatic binding energies is very small (**Table S1**).

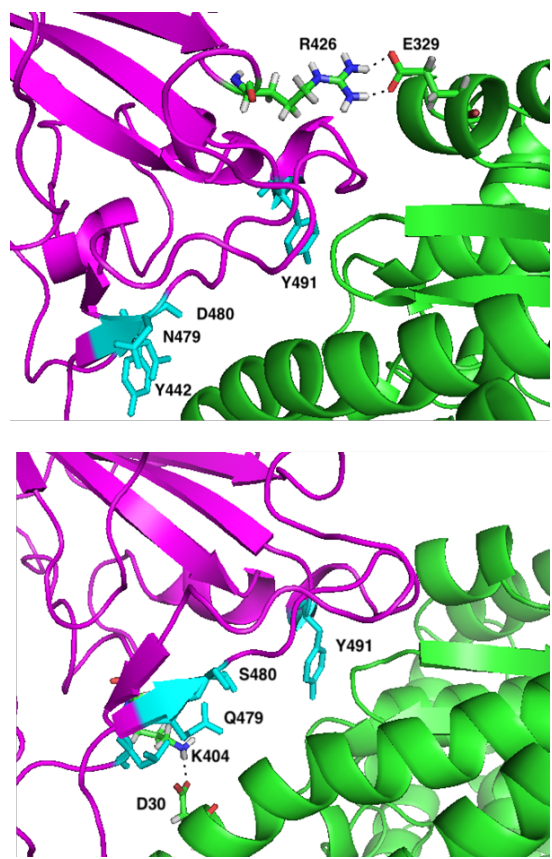

**Figure 2.** The MD optimized structure of SARS-CoV (top) and SARS-CoV-2 (bottom). The secondary structures of the spike protein and the ACE2 are shown in magenta and green respectively. Key residues are labeled and shown in cyan. The salt-bridges formed between R426 and E329 in SARS-CoV and K404 and D30 in SARS-CoV-2 are shown in black dots.

**Table 1.** Total Electrostatic and van der Waals interactions between SARS-CoV, SARS-CoV-2 and ACE2

| CoV-> ACE2 | CoV-2-> ACE2 |
|---|---|
| **Electrostatics** (Kcal/mol) | |
| R426 -> E329  -5.83 | K404 -> D30  -7.22 |
| K390 -> D37   -1.01 | Y491 -> E37  -1.28 |
| D393 -> K353  -0.95 | K390 -> D37  -1.21 |
| **Van der Waals** (Kcal/mol) | |
| Y475 -> T27  -1.56 | D491 -> K353  -2.10 |
| I489 -> Q325  -1.34 | Y475 -> K31  -1.50 |
| The major electrostatic contributions are observed for the R426/E329 salt-bridge for SARS-CoV and K404/D30 for SARS-CoV-2. Table S1 includes more information about the electrostatic and van derWaalsinteractions between SARS-CoV, SARS-CoV-2 and ACE2 | |

Furthermore, these mutants induce structural changes that increase the favorable van der Waals interactions in SARS-CoV-2. The maximum van der Waals attractions of -1.6 Kcal/mol are observed between CoV-Y457 and ACE2-T27 for SARS-CoV, while for SARS-CoV-2 the maximum attraction is -2.1 Kcal/mol observed between CoV2-D491 and CoV2-K353 (**Table 1**). The total van der Waal's contribution to the binding energy in case of SARS-CoV-2 is higher than SARS-CoV by ~ 4Kcal/mol. In total, the electrostatic and van derWaalsattraction between the SARS-CoV-2 and the ACE2 is stronger by ~7 Kcal/mol than the SAR-CoV.

Several studies proposed that increased virulence of SARS-CoV-2 is due to the increased binding affinity to the ACE2 receptor.[3,6,7] Yan et al., proposed that mutation $V404 \rightarrow K417$ may result in higher binding affinity due to the salt bridge between K417 and D30, while the $R426 \rightarrow N439$ mutation weaken the interaction with E329. However, our simulations show that the $V404 \rightarrow K417$ mutation increases the binding affinities by favorable electrostatic energies more than the energy loss induced by $R426 \rightarrow N439$ mutation. Additionally, mutation $L472 \rightarrow$ $F486$ was proposed to weaken van der Waals contact with Met 82 of ACE2. On the contrary, the relaxed structure shows a van derWaalsattraction of -1.32 Kcal/mol.[7] The structural changes induced by these mutations likely confer stability to CoV-2 as it binds to ACE2. This agrees with the study of Ortega et al.,[8] where they found that two main residues (479 and 487) have been associated with human ACE2 recognition. They further found a higher number of residues in the CoV-2 capping loops and attributed the more favorable binding affinity of -15.7 Kcal/mol as opposed to -14.1 Kcal/mol for CoV.[8]

Based on our calculations, there is no significant contribution from a single mutant to the binding energies between the SARS and the ACE2. However, these mutations induce sophisticated structural changes that enhance the electrostatic and van der Waals binding energies (Table S1). These results might also support the idea that it is unlikely that the SARS-CoV-2 is lab engineered but rather a result of a biological evolution.[18]

**AUTHOR INFORMATION**
MKS and AK worked on collecting the recent work published related to the CoV-19 to find out the relevant differences between the CoV-1 and CoV-2, in order to define the mutations. MA ran the MC and MD simulations.

**Notes**
The authors declare no competing financial interests.

**ACKNOWLEDGMENT**

This work is supported by the University of Groningen, and the Nanotechnology Research Centre (NTRC), the British University in Egypt (BUE). We would like to acknowledge the funding from NSF MCB 1519640 grant to MRG. We would like to thank Prof. Marilyn Gunner and Jochen Kupper for the useful discussion.

# Supporting Information for:
# Comparing the binding interactions in the receptor binding domains of SARS-CoV-2 and SARS-CoV

Muhamed Amin[†,‡]*, Mariam K. Sorour[§], Amal Kasry[§]

[†] Department of Sciences, University College Groningen, University of Groningen, Hoendiepskade 23/24, 9718 BG Groningen, The Netherlands
[‡] Center for Free-Electron Laser Science, Deutsches Elektronen-Synchrotron DESY, Notkestrasse 85, 22607 Hamburg, Germany
[§] Nanotechnology Research Centre (NTRC), the British University in Egypt (BUE), El Sherouk City, Suez Desert Road, Cairo 1183.

m.a.a.amin@rug.nl

**Supporting Information includes:**

1- The table of the electrostatic interactions of the amino acids in the receptor binding domain (RBD)

2- The constructed SARS-CoV-2 structure by mutating and optimizing SARS-CoV.

**Table S1.** The electrostatic interactions between amino acids in the RBD for SARS-CoV and SARS-CoV-2.

| ARS-CoV-2 | | | | | | SARS-CoV | | | | | |
|---|---|---|---|---|---|---|---|---|---|---|---|
| Van der Waals Interactions | | | Electrostatic Interactions | | | Van der Waals Interactions | | | Electrostatic Interactions | | |
| CoV-2 | ACE2 | Force (Kcal/mol) | CoV-2 | ACE2 | Force (Kcal/mol) | CoV | ACE2 | Force (Kcal/mol) | CoV | ACE2 | Force (Kcal/mol) |
| Y 491 | K 353 | **-2.1** | K404 | D 30 | **-7.22** | Y475 | T27 | **-1.56** | R 426 | E329 | **-5.83** |
| Y 475 | K 31 | **-1.5** | Y 491 | E 37 | **-1.28** | I 489 | Q 325 | **-1.34** | K 390 | E 37 | **-1.01** |
| Q 484 | K 353 | **-1.237** | K 404 | HIS 34 | **-1.24** | LEU 472 | M 82 | **-1.32** | D 393 | K 353 | **-0.95** |
| P 462 | T 27 | **-1.222** | K 390 | E 37 | **-1.21** | Y 442 | K 31 | **-1.24** | N 473 | Y 83 | **-0.73** |
| Y 475 | T 27 | **-1.207** | N 473 | Y 83 | **-0.99** | Y 475 | K 31 | **-1.2** | R 426 | D 329 | **-0.7** |
| F 472 | M 82 | **-1.102** | T 486 | Y 41 | **-0.96** | P 462 | T 27 | **-1.12** | D 480 | K 37 | **-0.64** |
| Q 484 | Y 41 | **-1.003** | K 404 | E 37 | **-0.67** | Y 484 | Y 41 | **-1.021** | K 390 | D 353 | **-0.63** |
| N 487 | K 353 | **-0.986** | K 390 | D 30 | **-0.61** | Y 491 | K 353 | **-0.993** | T 486 | Y 83 | **-0.59** |
| N 473 | Q 24 | **-0.973** | D 393 | R 393 | **-0.55** | T 487 | Y 41 | **-0.962** | Q 492 | E 329 | **-0.58** |
| I 489 | Q 325 | **-0.964** | D 393 | K 353 | **-0.55** | Y 491 | K 353 | **-0.892** | D 393 | R 37 | **-0.54** |
| V 498 | V 104 | **0.0** | D 393 | D 30 | **0.61** | V 498 | V 604 | **0.0** | D 480 | E 353 | **0.61** |
| V 498 | V 581 | **0.0** | D 392 | E 37 | **0.6** | V 498 | V 581 | **0.0** | D 392 | E 38 | **0.58** |
| V 498 | V 574 | **0.0** | K 404 | R 393 | **0.52** | V 498 | V 574 | **0.0** | D 393 | D 37 | **0.5** |
| V 498 | V 573 | **0.0** | Q 484 | K 353 | **0.5** | V 498 | V 573 | **0.0** | D 393 | D 393 | **0.48** |
| V 498 | V 506 | **0.0** | K 404 | K 353 | **0.46** | V 498 | V 506 | **0.0** | K 390 | HIS 353 | **0.46** |
| N 473 | Y 83 | **0.0005** | D 393 | E 37 | **0.73** | V 498 | V 491 | **0.0** | R 426 | R 38 | **0.42** |
| Y 491 | E 37 | **0 .025** | Q 484 | D 355 | **0.74** | T 486 | Y 41 | **0.001** | K 390 | R 393 | **0.71** |
| N 473 | Y 83 | **0.1515** | Q 484 | Y 41 | **0.78** | T 486 | Y 41 | **0.482** | D 480 | D 38 | **0.92** |

| Y 491 | E 37 | 0.324  | K 390 | R 393 | 0.81   | T 486 | Y 41  | 0.482  | D 393 | E 37  | 0.73   |
| K 404 | D 30 | 0.637  | K 390 | K 353 | 1.02   | R 426 | E 329 | 3.446  | K 390 | K353  | 1.24   |
|       |      | -11.14 |       |       | -8.51  |       |       | -7.242 |       |       | -5.550 |